\documentstyle[epsf]{aipproc}

\def\eg{{\it e.g.\ }}
\def\ltsima{$\; \buildrel < \over \sim \;$} 
\def\simlt{\lower.5ex\hbox{\ltsima}}
\def\gtsima{$\; \buildrel > \over \sim \;$} 
\def\simgt{\lower.5ex\hbox{\gtsima}}

\title{QUANTITATIVE MORPHOLOGY AT HIGH REDSHIFTS}
\author{Roberto G. Abraham}
\address{Royal Greenwich Observatory, Madingley Road, Cambridge CB3 OEZ, UK}

\begin{document}
\maketitle  

\begin{abstract}
The current evidence for morphologically peculiar galaxy populations
at high-redshifts is outlined. After describing various techniques
which can be used to quantify the importance of ``morphological
K-corrections'', and to objectively classify galaxy morphology in the
presence of these biases, it is concluded that observational biases
are not sufficient to explain the increase in the fraction of peculiar
galaxies on deep HST images. A new technique is then described which
models the spatially resolved internal colors of high redshift
galaxies, as a probe of the processes driving galaxy evolution. This
``morphophotometric'' approach investigates directly the evolutionary
history of stellar populations, and is a sensitive test of the
mechanisms through which galaxies build up and evolve in the field. As
a case study, we analyse several ``chain galaxies'' in the Hubble Deep
Field. These chain galaxies are shown to be protogalaxies undergoing
their first significant episodes of star-formation, and not simply
distant edge-on spirals.
\end{abstract}

\section{Introduction}

Recent work from deep imaging
\cite{Griffiths:1994,Glazebrook:1995,Driver:1995,Abraham:1996a,Abraham:1996b,Giavalisco:1996}
and spectroscopic \cite{Lilly:1995,Cowie:1995,Lilly:1996,Ellis:1996}
surveys has shown that much of the rapidly evolving faint blue galaxy
population \cite{Broadhurst:1988,Koo:1992,Lilly:1995} is comprised of
morphologically peculiar galaxies. These systems may be luminous
counterparts to local irregular galaxies, tidally disturbed systems,
or perhaps members of entirely new classes of objects with no local
counterpart. Another possibility is that these morphologically
peculiar systems are simply ``ordinary'' galaxies whose strange
appearance is simply a result of their being observed in the
rest-frame ultraviolet (a ``morphological K-correction''), where we
know little about the appearance of the galaxy population. This
distinction between intrinsic and apparent peculiar galaxies lies at
the heart of this meeting. In this article several lines of evidence
are reviewed which suggest that the bulk of the morphological
peculiarities seen in distant galaxies are intrinsic to these systems,
and not simply the product of rest-frame bandshifting. In the final
section of this article preliminary results from a new line of
evidence are presented, focussing on the ``chain galaxy'' population
as a case study.

\section{High Redshift Galaxy Morphology}

\subsection{Quantitative Comparisons with Simulations}

The first line of evidence is the result of a quantitative comparison
between the observed morphologies of local galaxies and the predicted
appearance of their {\em non-evolved} high-redshift counterparts. In
order to undertake this comparison we have developed a technique for
artificially redshifting local galaxy CCD images by assigning separate
spectral energy distributions to individual pixels. Spectral energy
distributions for each pixel are determined by using optical colors to
interpolate between template spectra corresponding to local S0, Sab,
Sbc, Scd, Sdm, and starbursting galaxies.  Figure~\ref{fig:artz}
illustrates the power of this technique by showing the excellent
agreement between ``predicted'' and observed far-UV morphologies for a
typical galaxy in our calibration sample\footnote{In order to make a
comparison with UIT straightforward, noise has not been added to the
simulations. Because of $(1+z)^4$ cosmological dimming extremely long
exposure times with HST would be required to image this galaxy at high
redshifts with a signal-to-noise level equivalent to that
shown.}. This figure illustrates the extreme case in which the
$I$-band morphology is extrapolated to the far-UV, corresponding
to redshifts $z>4$. In fact, at $z>4$ the galaxian internal dynamical
timescale is a substantial fraction of the age of the Universe, so the
existence of morphologically unevolved galaxies is not expected.

\begin{figure}
	\epsfxsize=3.0in 
        \begin{center}
	\hspace{0.3in}
        \end{center}
	\caption[UIT]{Comparison between the far-UV morphology
	``predicted'' on the basis of {\em optical} colors of NGC1365
	and the observed morphology from the Ultraviolet Imaging
	Telescope (UIT). Optical images were obtained from the 2.5m du
	Pont telescope on Las Campanas, and no noise has been added in
	order to match the signal-to-noise characteristics of the
	UIT.}  \label{fig:artz}
\end{figure}

\begin{figure}[htp]
	\begin{tabular}[t]{cc} 
		\begin{minipage}[t]{3.0in}
			\epsfxsize=2.8in 
			\epsfbox{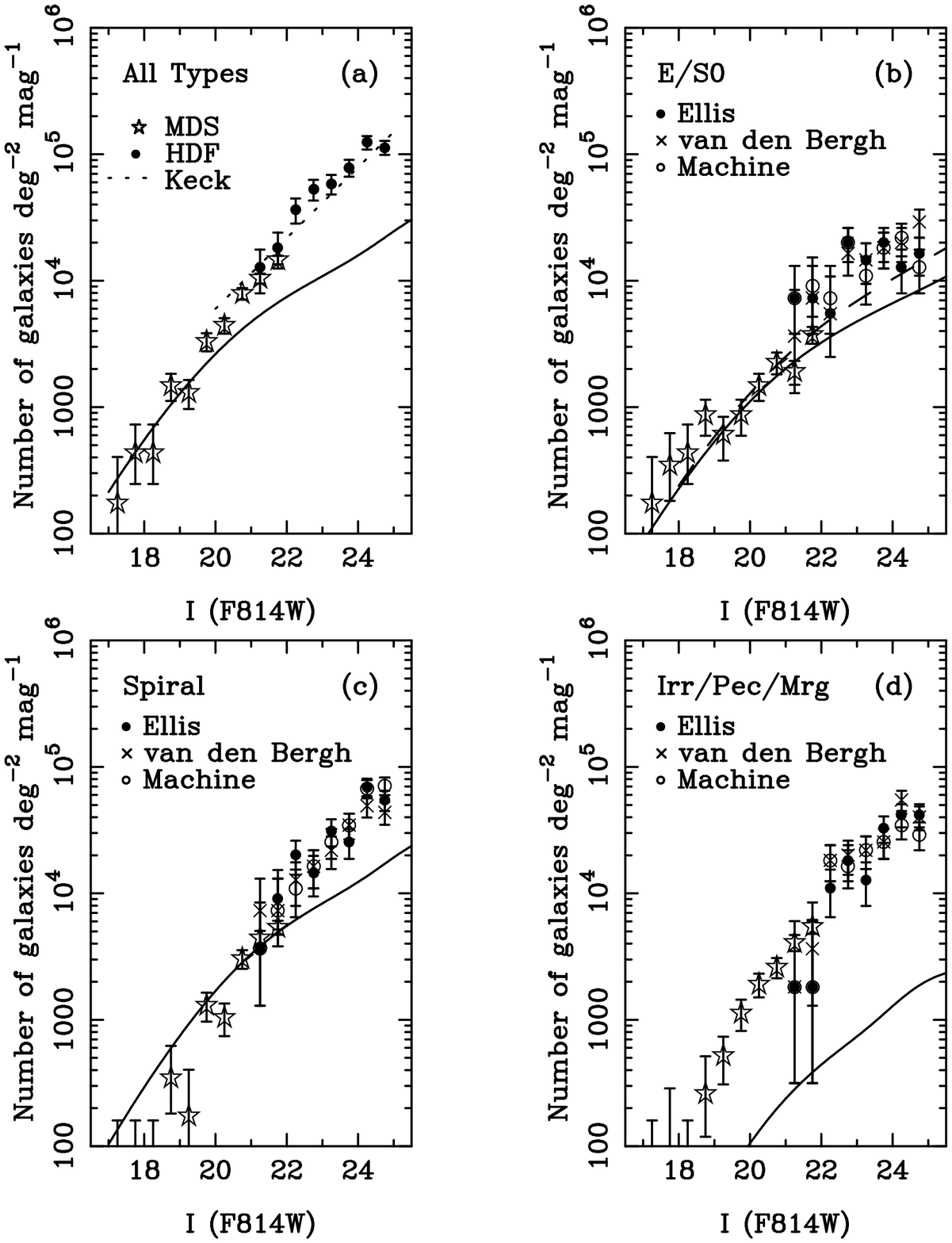} 
			\caption[Number Counts]{The number-magnitude 
             relations for morphologically segregated samples of galaxies 
             from the HDF and MDS (from Abraham {\em et al.} 1996a). Open
             circles indicate counts obtained from automated
             classifications, closed circles
             indicate the results from the visual classifications of Ellis,
             and crosses indicate the results from the visual
             classifications of van den Bergh. The MDS counts are indicated by the
             stars on each panel. The no-evolution $\Omega=1$ curves from
             Glazebrook~{\em et~al.}~(1995), extrapolated to $I=25 {\rm
             ~mag}$, are superposed. The dashed line on the E/S0 diagram
             shows the effect of assuming $\Omega=0.1$. The dotted line
             in panel (a) shows the $I$-band number counts determined by
             Smail~{\em et al.} (1995) from two deep fields imaged with
             the Keck telescope.}  
        \label{fig:counts} 
		\end{minipage} & 
		\begin{minipage}[t]{3.0in}
			\epsfxsize=2.8in
			\epsfbox{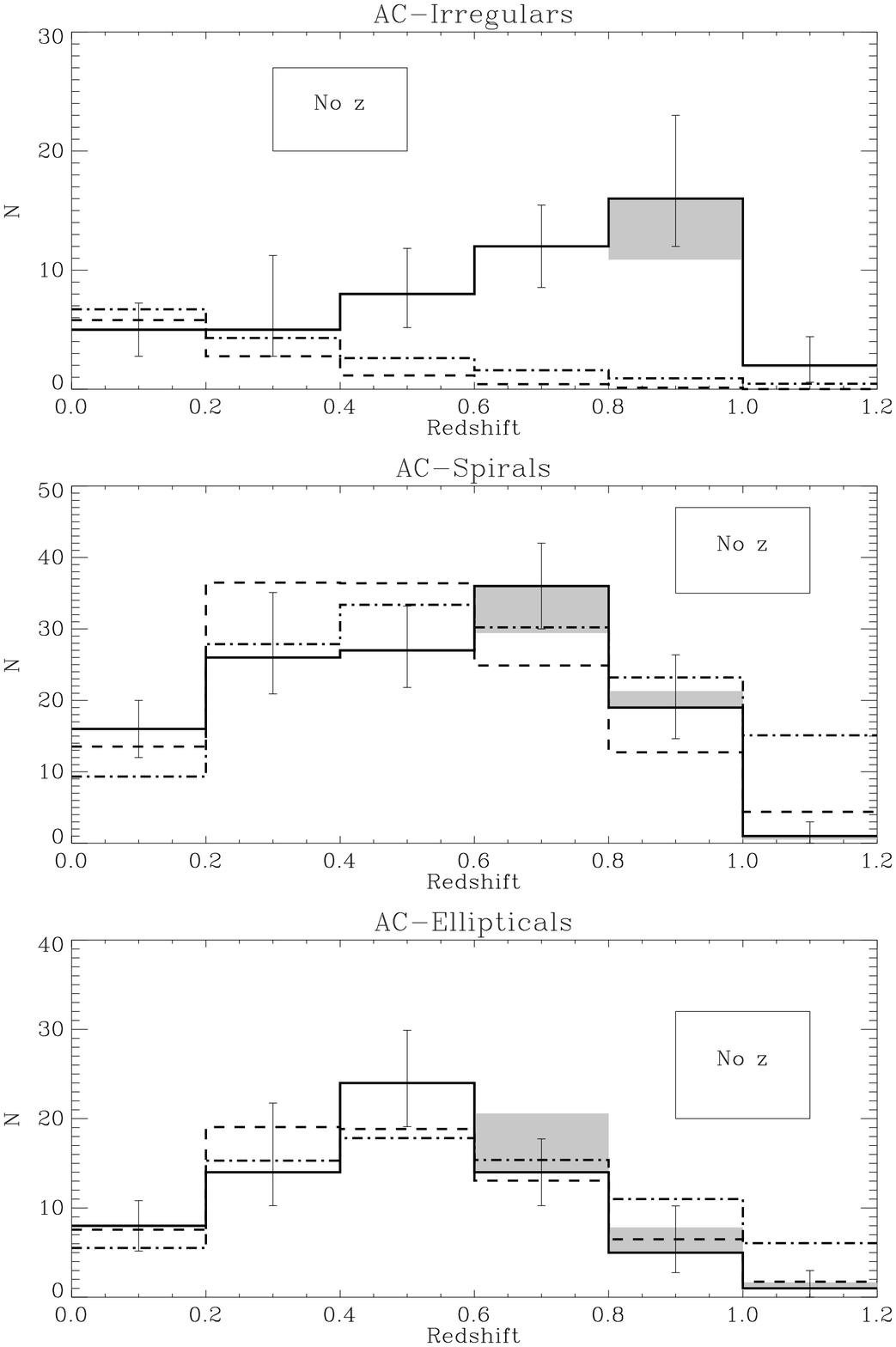} 
			\caption[NZ]{Morphologically segregated number counts from
			Brinchmann {\em et al.} 1997 (in preparation), based on data
			from the CFRS/LDSS collaboration. The bins show counts as a 
			function of redshift for irregular/peculiar/merger systems 
			(top), spirals (middle),and ellipticals (bottom). Morphological
			classifications
			have been made from WF/PC2 images using the central concentration 
			vs. asymmetry
			system described in the text and in Abraham {\em et al.} 1996a.
			The shaded region corresponds to the size of the ``morphological 
			K-correction'' on the classification. Superposed on the
			histograms are the predictions of no-evolution and 1 mag
			linear evolution to $z=1$ models. }
			\label{fig:nz}
		\end{minipage}
	\end{tabular}
\end{figure}

With this capability to model the effects of bandshifting on observed
morphology, {\em and by assuming an approximate redshift
distribution}, one is able account for the impact of morphological
K-corrections on morphologically segregated deep number-magnitude
counts. Because the subjective nature of visual classification makes
comparisons between different groups susceptible to large systematic
errors \cite{Naim:1995}, the best approach to making these counts is
to adopt a quantitative morphological classification system. Several
quantitative classification systems have now been
developed\cite{Abraham:1994,Abraham:1996b,Odewahn:1996}.
Figure~\ref{fig:counts} shows the number-magnitude counts which result
from applying a particularly simple system (based on measurements of
central concentration, $C$, and asymmetry, $A$) to data from the
Hubble Deep Field. Also shown are the no-evolution predictions for
ellipticals, spirals, and irregular/peculiar/merger systems,
constructed as described in Glazebrook~{\em et~al.}~(1995) and
Abraham~{\em et~al.}~(1996a,b), by adopting Schechter luminosity
functions (LFs) with parameters given by Loveday {\em et al.} (1992),
and a high normalization $\phi_\star=0.03 h^3$ Mpc$^{-3}$.  The
predicted counts for the elliptical galaxies are based on a flat slope
($\alpha=-1$) for the faint-end of the LF, rather than the turn-over
originally found by Loveday {\em et al}. The steep counts for the
irregular/peculiar/merger systems continues to the limits of the
survey. Beyond $I_{814}=22 {\rm ~mag}$ the spiral counts show a
significant excess over the no-evolution predictions. A weaker trend
is seen for the spheroidal systems (whose counts are only marginally
above the no-evolution prediction) and there is some evidence of a
turn-over in the last magnitude interval.

\subsection{Redshift Surveys}

The analysis presented in the previous section is fairly sensitive to
the assumed redshift distribution of the peculiar galaxy
population. However recent evidence from redshift surveys, most
notably the Canada-France Redshift Survey (CFRS) suggests that, at
least to $I<22$ mag (the typical morphologically resolved magnitude
limit for deep HST imaging, prior to the Hubble Deep Field), the great
majority of objects are at redshifts $z<1.5$
\cite{Lilly:1996}. Therefore even quite deep HST WF/PC2 $I_{814}$-band
(the band most commonly used to quantify the morphological composition
of the distant field) images should generally be compared with
$B$-band or $U$-band local galaxy data.  Since most local surveys of
galaxy morphology are based on blue-sensitive photographic plates, the
rest-frame $B$-band is in fact where we are most familiar with the
appearance of galaxies, and the effects of bandshifting on these
galaxies are likely to be less significant than the effects of limited
signal-to-noise or binning. Images of local galaxies in $U$-band
(blueward of the 4000\AA~break) are often substantially more irregular
than $B$-band data, but they do not yet show the fantastic
morphological variations (such as disappearing bulges) seen in images
in the far-UV ($\sim1500$\AA), as presented by O'Connell and others at
this conference.

This line of reasoning has recently been spectacularly confirmed
\cite{Brinchmann:1997} by Brinchmann {\em et al.} (1997), who have applied
an objective classification scheme, again calibrated using
pixel-by-pixel K-corrections, to a set of $\sim 300$ HST
$I_{814}$-band images of galaxies {\em with known redshifts} taken
from the CFRS and LDSS \cite{Ellis:1996} surveys. Because the
statistical completeness of this sample is very well understood,
reliable number-redshift histograms can be constructed for the various
morphological types. The morphologically resolved $n(z)$ result
obtained by Brinchmann {\em et al.} is shown in Figure~\ref{fig:nz},
and confirms that irregular/peculiar/merging systems are already
greatly in excess of the predictions of no-evolution and
mild-evolution models at redshifts $z\sim1$.  It is emphasized that:
(a) because of the complete redshift information, morphological
K-corrections have been accounted for explicitly for each galaxy in
this study, and (b) in any case morphological K-correction effects
cannot be dominant in this survey, because the peculiar excess is
already large by $z=1$, at which point one is only just beginning to
probe into the ultraviolet.

\section{Morphophotometry}

Studies of evolution in high-redshift galaxies have focused generally
on either analyses of morphological characteristics, or on analyses of
{\em integrated} colors using spectral synthesis techniques. While
morphological studies often point the way forward, only rarely do they
probe the underlying physics of galaxy evolution. On the other hand,
studies of integrated colors are also limited, because galaxies are
not homogeneous systems.  Galaxy evolution can be described as a
series of punctuated star-formation episodes whose imprints are
recorded in distinct stellar populations (such as the disk and bulge),
so studies which resolve these stellar populations are required in
order to understand the mechanisms through which galaxies are built-up
and evolve.

Therefore it seems that the important next step in understanding the
history of galaxies is to unify morphological studies with stellar
evolution modelling. In this section we present preliminary results
from just such a {\em morphophotometric} analysis of the ``chain
galaxy'' population. We have chosen to focus on this subset of the
high-redshift peculiar galaxy population in this article because the
controversy surrounding these systems is a perfect encapsulation of
the theme of this meeting. In their discovery paper, Cowie and
collaborators \cite{Cowie:1995} claimed to have found a significant
new young population of galaxies at high redshifts. This claim was
immediately disputed \cite{Dalcanton:1996}. Dalcanton \& Shectman
(1996) claim that chain galaxies are simply the distant counterparts
to local edge-on low surface-brightness spiral galaxies. Can the
internal colors of these systems shed light on the nature of these
objects?

\subsection{Resolved Color Modelling}

Consider the central equation of population synthesis
\cite{Bruzual:1993}, commonly used to model spectral evolution in
galaxies:

\begin{equation}
F_\lambda(T) = \int_{0}^{T} \Psi(T-\tau) f_\lambda(\tau) d\tau.
\label{eq:synthesis}
\end{equation}

In this formulation the emergent flux from a galaxy, $F_\lambda$, at
time $T$ is described by the convolution of the spectrum of an
evolving instantaneous starburst, $f_\lambda(t)$, with an assumed
star-formation rate (SFR) function $\Psi(t)$. The $f_\lambda(t)$ term
is in principle known for various choices of initial mass function
(IMF) and metallicity, using libraries of template stellar spectra and
isochrones. Hierarchical formation scenarios suggest that stellar
populations become built up over time, so in reality the age $T$ of a
galaxy is not a constant, and the distribution of internal colors is a
partial record of the formation timescale(s) of the system. This age
information for a given stellar population is diluted by the
convolution with $\Psi(t)$, whose form is usually unknown. But in
regions where the resolution is high enough to resolve the sites of
current star-formation, or where mixing of multiple generations of
stars has not occured, $\Psi(t)$ can be approximated by a
$\delta$-function, breaking the convolution degeneracy in Equation 1,
and allowing direct measurement the age distribution and form of
$f_\lambda(t)$.  Consider, for example, the canonical picture of a
late-type spiral galaxy, where the bulk $\Psi(t)$ can be
well-approximated by a constant star-formation rate.  This
``constant'' overall star formation rate is physically simply a
time-average over the appearance and disappearance of spatially
distinct HII regions and star-formation complexes, each of which
individually can be considered to be a bursting simple stellar
population with a lifetime (before disruption or gas depletion) that
is short compared to the dynamical timescale of the galaxy.  Therefore
the distribution of colors for individual resolved young stellar
associations on a color-color diagram directly maps out the shape of
$f_\lambda(t)$ for a set of young ages, giving direct access to the
integrand of Equation~1 without first filtering by a convolution. As
these stellar associations age and disappear the convolution with
$\Psi(t)$ in Equation 1 becomes important, as young stars become
assimilated into older galactic components (ie. the disk and bulge)
and are spatially averaged with earlier generations of stars. The
distribution of colors for older stellar populations (\simgt 1 Gyr)
are therefore expected to trace out a continuous age track on the
color-color diagram. The {\em shape} of this track for older stellar
populations is in effect a measurement of the form of the
star-formation law $\Psi(t)$, while the {\em distribution} of colors
along this track is a record of the uniformity with which episodes of
star formation have added to the stellar population (\eg via numerous
small bursts, or a smaller number of larger bursts).

\begin{figure}
	\epsfxsize=4.0in \begin{center} \hspace{0.15in}
	\epsfbox{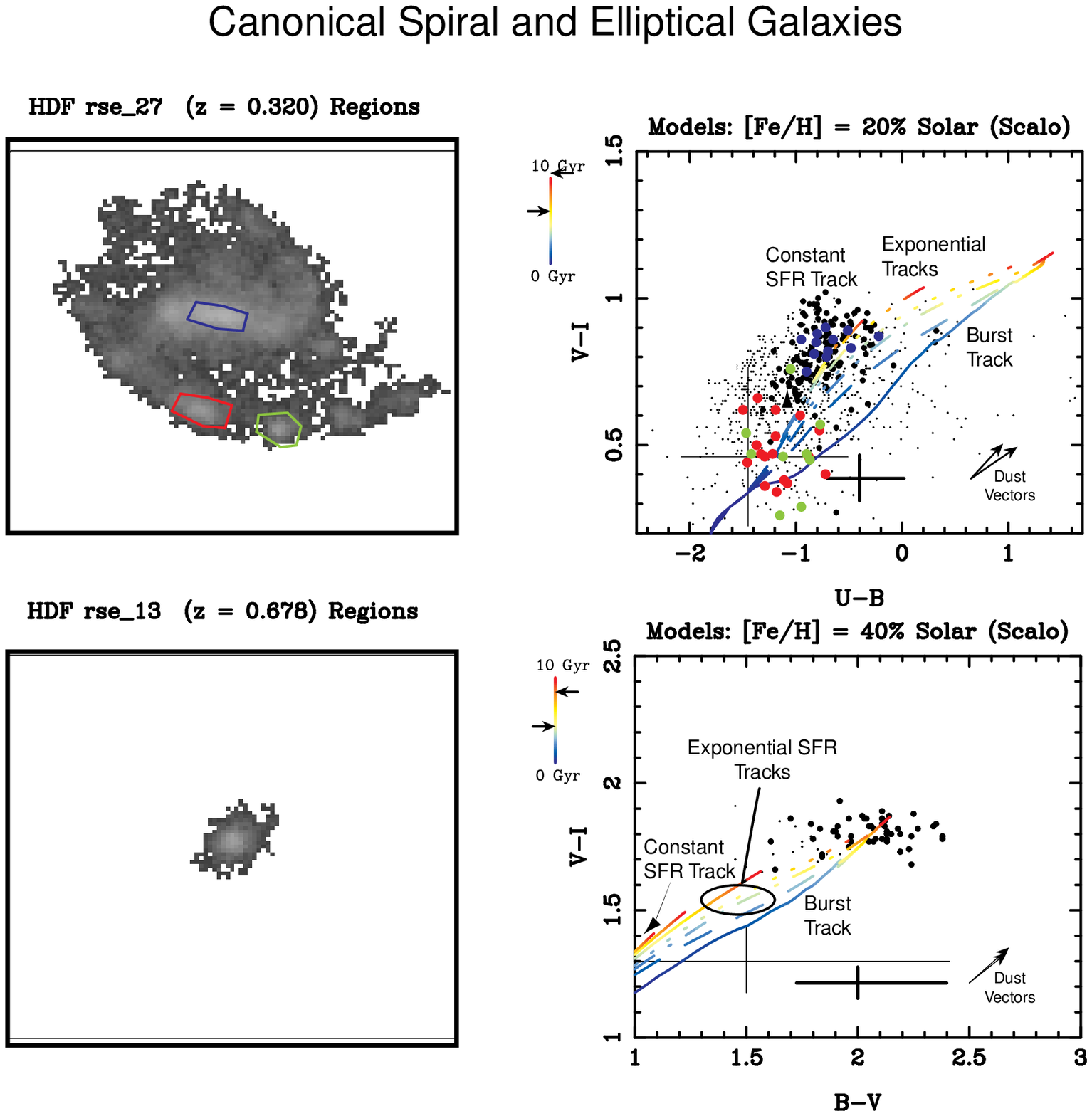} \end{center}
	\caption[UIT]{Morphophotometric color-color diagrams for two
	``canonical'' galaxies in the Hubble Deep Field. The panels on
	the left show $I$-band images of the galaxies, ``segmented''
	from the background sky by isophotal thresholding. The right
	hand panels show the color-color diagrams for {\em individual
	pixels} in the galaxy. The data points have been subdivided
	into high signal-to-noise (SNR) pixels [circles] and low
	signal-to-noise pixels [dots]. The mean error bars for the
	high SNR and low SNR points are shown by the dark and light
	error bars. Model tracks are also shown on the right, and are
	calibrated to age in Gyr [keyed to the color bar]. Arrows on
	the color bar indicate the age of the Universe in the rest
	frame of the galaxy, for $H_o=70$~km/s/Mpc and $\Omega=0.1$
	and $\Omega=1$. The dust vectors (for LMC and SMC extinction
	laws) shown correspond to an extinction of $A_B=0.2$ mag in
	the rest frame of the galaxy.  The colored points shown on the
	color-color diagram for the spiral galaxy correspond to pixels
	inside the colored polygons shown on the left hand panel.}
	\label{fig:spirell}
\end{figure}

Figure~\ref{fig:spirell} shows the resolved color-color diagram for
two rather typical spiral and elliptical galaxies in the HDF. These
exhibit star-formation characteristics that are in reasonable
agreement with our expectations based on studies of stellar
populations in local galaxies. For example, the spiral galaxy is well
described by a roughly constant star formation history. The bulge is
the oldest component of this system, has a small dispersion in color,
and is several Gyr older than the disk. By contrast the elliptical
system shown is well described by an exponential star-formation
history with a short e-folding timescale (around $\tau = 1$ Gyr.) This
system is apparently rather old and (from the small dispersion in
color), all parts of the galaxy are well mixed and roughly coeval. The
majority of spirals and ellipticals in the Hubble Deep Field exhibit
morphophotometric diagrams qualitatively similar to those shown in
Figure~\ref{fig:spirell}.

\begin{figure}
     \epsfxsize=5.0in \begin{center} \hspace{0.25in}
     \epsfbox{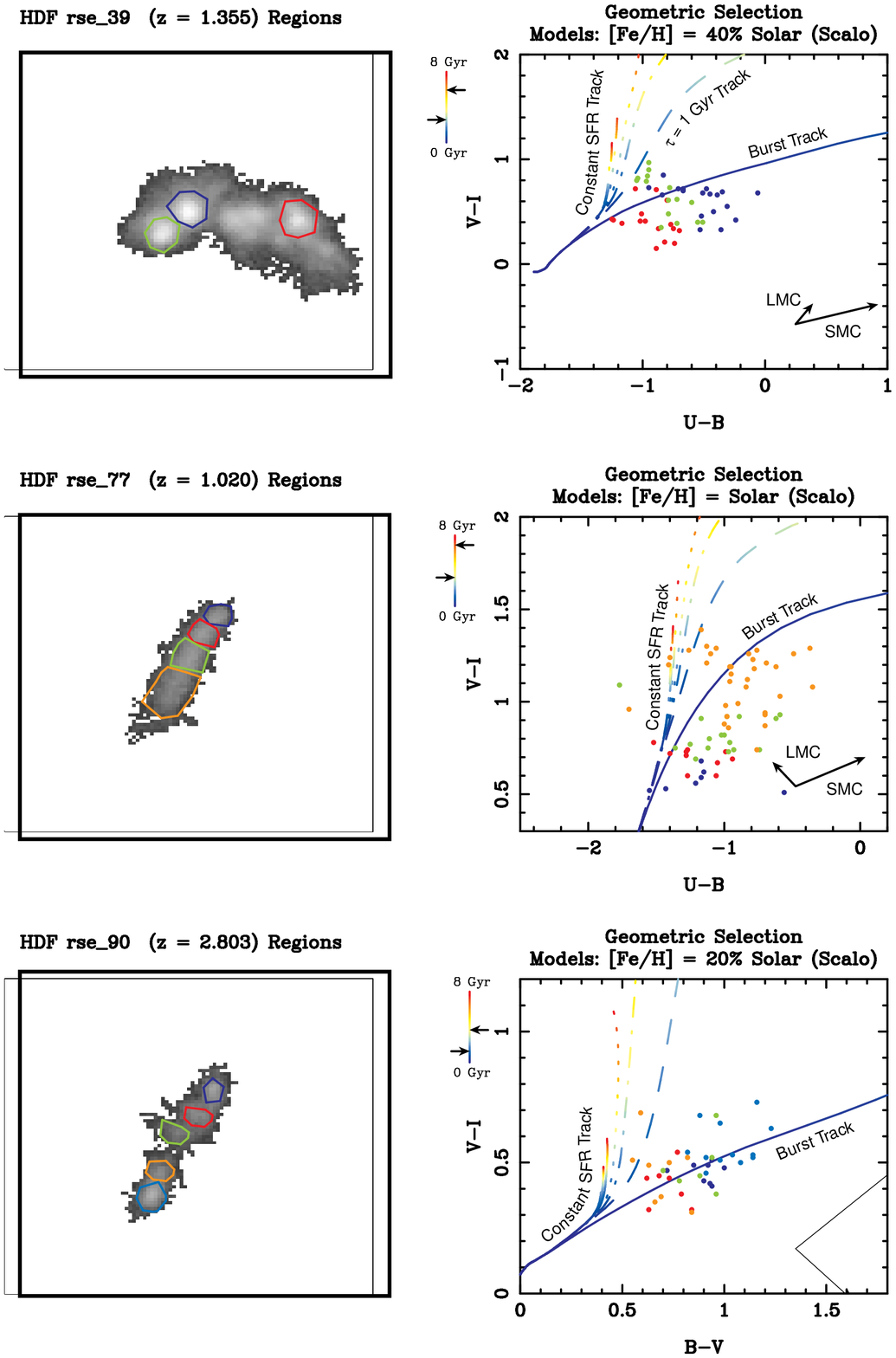} \end{center} \caption[UIT]{Morphophotometric
     color-color diagrams for the three ``chain galaxies'' in the
     public database of HDF redshifts. The right-hand panels show the
     distribution of pixel colors within the regions of the galaxy
     defined on the left-hand panels. Also shown on the right are models
     corresponding to constant star formation, exponential star
     formation, and an instantaneous starburst. See also the caption
     for the previous figure.}  \label{fig:chain}
\end{figure}

\subsection{The Nature of Chain Galaxies}

Figure~\ref{fig:chain} shows a montage of resolved color-color
diagrams for the three ``chain galaxies'' (as visually classified by
R. Ellis) with known redshifts in the Hubble Deep Field\footnote{HDF
redshifts discussed in this article were taken from the 55 redshifts
available on the World Wide Web as of April 1997.}. The
morphophotometric analysis shown in this figure indicates that these
systems are likely to be very young galaxies recovering from their
first star-formation episodes. Unlike the case for late-type galaxies
exhibiting knots of star-formation superposed on a disk (as shown in
Figure~\ref{fig:spirell}), there is no evidence for an underlying
``old'' component in any of these systems. Using the surface
brightness detection threshold for the $I_{814}$ data, we can place an
upper limit of $<10\%$ for the contribution of old stars to the total
baryonic mass of two systems at $z<1.5$ systems shown in this
figure\footnote{Strong upper limits cannot be placed on a putative old
component underlying the ``hot dog'' Lyman limit system at $z=2.803$
shown in Fig.~\ref{fig:chain}, because of strong K-corrections for
red light at $z>2$. However, depending on $\Omega$, restrictions on
such a component may be imposed by the age of the Universe in the rest
frame.}. All components of the chain galaxies lie close to the
``pure'' starburst track: evidently rather little mixing of young
components within the body of the galaxy as occurred. (This mixing
would show up as a dispersion along the exponential or constant star
formation tracks, as populations with different ages blend
together). Intriguingly, the knots of star-formation in the
lower-redshift chains appear to be synchronized, both spatially and
temporally.  Star-formation has been triggered along the body of these
system like a string of fireworks. The oldest knots in the $z<1.5$
systems appear to have ignited the other knots in sequence along the
body of the galaxies. In all cases shown the unweighted mean age of
the starlight in the galaxy is 100-200 Myr (comparable with the
dynamical timescale of the galaxy), with the youngest and oldest knots
in the chain differing in age by around 30-50 Myr. {\em This
morphophotometric analysis indicates that these chain galaxies are
likely to be stochastically ignited very young galaxies, and not
edge-on low-surface brightness spirals, as has been claimed.}
Although it is important to bear in mind that the three systems
presented in this article may not be representative of the class
(being taken from an incomplete redshift survey, with strong biases
toward strong emission-line systems), it appears that at least {\em
some} chain galaxies ({\em i.e.}~all chain galaxies in the redshift
sample to date) are protogalactic starburst systems.

\section*{Conclusions}

The taxonomy of peculiar galaxies is confusing and possibly
inconsistent, but studies using objective classifications calibrated
by simulations offer a way forward. These studies show that
contamination by bandshifted late-type galaxies into samples of
``peculiar'' galaxies is small until $z \sim 1.5$, at which point the
observed fraction of irregular/peculiar/merging systems is already
much higher than predicted on the basis of no-evolution models. We
therefore conclude that ``morphological K-corrections'' are a
second-order effect, and do not account for the proportion of
irregular/peculiar/merging systems seen on deep HST images. At higher
redshifts contamination by bandshifted spirals may be more important,
although genuinely protogalactic systems are definitely seen. At least
some ``chain galaxies'' are protogalaxies -- strongly starbursting
systems forming their first generation of stars -- and {\em not} not
edge-on low surface brightness systems.

\bigskip
\noindent{\bf Acknowledgments} I thank my collaborators Richard Ellis,
Jarle Brinchmann, Karl Glazebrook, Andy Fabian, Sidney van den Bergh,
Nial Tanvir, and Basilio Santiago for their many contributions to the
projects described in this article. I am also grateful to Simon Lilly
and the rest of the CFRS team for useful discussions, and for
permission to describe results in advance of publication.

\end{document}